\documentstyle[aps,preprint]{revtex}\tighten

\begin{document}

\def \D {\mbox{D}}
\def \d {\mbox{d}}
\def \t {\tilde}
\def\div {\mbox{div}\,}
\def\rd {\displaystyle{\cdot}}
\def\c {\mbox{curl}\,}
\def\ep {\varepsilon}
\def \ub {u_{\!_B}}
\def \vb {v_{\!_B}}
\def \eb {e_{\!_B}}
\def \st {\sigma_{\!_T}}
\def \ne {n_{\!_E}}
\def \ts {\textstyle}
\def\be {\begin{equation}}
\def\ee {\end{equation}}
\def\bea {\begin{eqnarray}}
\def\eea {\end{eqnarray}}
\def\la {\langle}
\def\ra {\rangle}
\def\p  {\partial}
\def\bi {\bibitem}
\def\case#1/#2{\textstyle\frac{#1}{#2} }

\title{Nonlinear Effects in the
Cosmic Microwave Background
\thanks{Invited contribution to Relativistic Cosmology, a
symposium celebrating GFR Ellis' 60th, Cape Town, February 1999}
}

\author{Roy Maartens\thanks{roy.maartens@port.ac.uk}}

\address{
Relativity and Cosmology Group, Division of Mathematics
and Statistics,\\ Portsmouth University, Portsmouth~PO1~2EG,
England.}

\maketitle

\begin{quote}

{\small Major advances in the observation and theory of cosmic
microwave background anisotropies have opened up a new era in
cosmology. This has encouraged the hope that the fundamental
parameters of cosmology will be determined to high accuracy in the
near future. However, this optimism should not obscure the ongoing
need for theoretical developments that go beyond the highly
successful but simplified standard model. Such developments include
improvements in observational modelling (e.g. foregrounds,
non-Gaussian features), extensions and alternatives to the simplest
inflationary paradigm (e.g. non-adiabatic effects, defects), and
investigation of nonlinear effects. In addition to well known
nonlinear effects such as the Rees-Sciama and Ostriker-Vishniac
effects, further nonlinear effects have recently been identified.
These include a Rees-Sciama-type tensor effect, time-delay effects
of scalar and tensor lensing, nonlinear Thomson scattering effects
and a nonlinear shear effect. Some of the nonlinear effects and
their potential implications are discussed.
}
\end{quote}

\section*{1. Introduction}

In 1948, Alpher and Herman predicted a cosmic microwave background
(CMB) radiation, which remained a theoretical possibility until the
serendipitous discovery by Penzias and Wilson in 1965. A potential
crisis for cosmology, arising from the apparent perfect isotropy of
the CMB, was turned into a dramatic success when in 1992 the COBE
satellite detected small anisotropies at the level predicted by the
standard inflationary cold dark matter model. The confluence of
theoretical and observational successes has opened up a new era of
precision-tested cosmology, exemplified by the upcoming launch of
the MAP and Planck satellites (see e.g. \cite{bops}). Together with
parallel advances in galactic, supernovae, lensing and other
observations, this has promoted the hope that the fundamental
parameters of the standard cosmological model ($\Omega_{\rm m}$,
$\Omega_\Lambda$, $h$, $n$, \dots) will be determined to high
accuracy, and that in some sense, cosmology will be ``solved" up to
details of fine-tuning.

However, physics rarely turns out be as simple as this, and
optimism needs to be tempered by a realisation that future
observations could entail not only the resolution of old problems,
but also the generation of new and unexpected ones (see also
\cite{s}). The new ``golden age" of cosmology is perhaps better seen
as the end of the beginning, rather than the beginning of the end.

It is clearly necessary to refine and develop the predictions of
the standard model in readiness for the emerging observational
data, and much effort has been put into this (see e.g.
\cite{HS}--\cite{dk}).
 But it is also necessary to develop beyond the simplified
standard model, in recognition of the rich complexity of the
universe, and in anticipation of new surprises from future
observations. This encompasses a range of theoretical advances:
\begin{itemize}

\item The modelling of observations needs significant development in
relation to foreground extraction, non-Gaussian features, data
analysis, etc. (see e.g. \cite{tjhl,vwhk}).

\item Extensions and alternatives to the simple inflationary paradigm
need to be explored, including investigation of non-adiabatic
effects (during inflation and reheating), of string cosmology (see
e.g. \cite{mvdv}), and of defect theories (see e.g.
\cite{udr,chm}). Such exploration can lead to unexpected results;
for example, it has recently been shown \cite{bkm} that resonant
amplification of metric perturbations in preheating after inflation
could re-process the primordial power spectrum, and leave an
imprint on CMB anistropies.

\item
Nonlinear effects in CMB anisotropies require further
study.
\end{itemize}

This article focuses on the latter aspect, in particular on
recently identified nonlinear effects. Well-known nonlinear effects
include the Ostriker-Vishniac \cite{ov} and kinetic
Sunyaev-Zeldovich \cite{sz2} effects, which are local effects that
arise from the modulation of the Doppler effect with
inhomogeneities in the optical depth. These effects could restore
small-scale power via hot cluster gas or re-ionisation (see e.g.
\cite{hk}). The Rees-Sciama effect \cite{rs} arises as a
second-order contribution to the integrated Sachs-Wolfe term.
Nonlinear integrated effects can in principle be significant
because of the long photon paths after last scattering. In practice
the Rees-Sciama effect is typically very small. More recently, it
has been shown \cite{sel} that a larger effect follows from
gravitational lensing of CMB photons by density perturbations.

Considerable effort has been put into calculating the imprint of
these particular nonlinear effects on CMB anisotropies, and this
work is of great importance for isolating the secondary
anisotropies and revealing the primordial spectrum \cite{rsh}. What
has received less attention is a systematic approach to nonlinear
effects in general. This is important because it could reveal new
sources of secondary anisotropies that may be significant.
Furthermore, a systematic analysis of second-order effects forms
the basis for estimating the {\em theoretical} errors associated
with the linear analysis on which CMB science rests.

Mollerach and Matarrese \cite{mm2} recently developed a systematic
second-order Sachs-Wolfe analysis, building on previous work by
Pyne and Carroll \cite{pc}. They found the second-order metric
perturbations to a matter-dominated flat universe in the Poisson
gauge by transforming the known synchronous gauge solutions, using
the methods of Bruni et al. \cite{bmms}. Then these solutions were
used to find the second-order Sachs-Wolfe effect. The main relevant
effects which they identified are:
\begin{itemize}
\item
in addition to the Rees-Sciama effect, another
nonlinear integrated Sachs-Wolfe effect representing corrections
to the linear gravitational wave contribution;
\item
in addition to gravitational lensing by linear density
perturbations, gravitational lensing by linear gravitational wave
modes;
\item
time delay effects of scalar and tensor lensing;
\item
a coupling of the velocity at last scattering with the perturbed
photon wave vector.
\end{itemize}

An alternative approach to nonlinear effects is developed in
\cite{mge}, and will be briefly described here. Instead of starting
from a background model and perturbing, this approach starts from
the fully nonlinear model of the inhomogeneous universe. The
approach is thus more general, although this generality comes at
the cost of greater difficulty in quantifying the effects on CMB
anisotropies. More importantly, the alternative approach covers
nonlinearity not only in the metric, but also in Thomson
scattering, and in relative motions of the particle species.
(Second-order scattering effects have previously been investigated
via a different approach by Hu et al. \cite{hss}.) It thus offers a
foundation for a comprehensive second- and higher-order analysis,
taking into account all sources of nonlinearity. (Nonlinear
polarisation has not been treated, although key elements of such a
treatment can be found in Challinor's linear analysis \cite{c}.)

The key local effects identified in the new approach are:
\begin{itemize}
\item
at each multipole order $\ell$, a coupling of baryonic bulk
velocity to the radiation brightness multipoles of order
$\ell\pm1$;
\item
a coupling of the acceleration, vorticity and shear at order $\ell$
to the brightness multipoles of order $\ell$, $\ell\pm1$,
$\ell\pm2$;
\item
for $\ell\gg1$, the shear coupling is dominant and could be
important.
\end{itemize}

These local effects need to be integrated to evaluate their impact
on CMB anisotropies, but a qualitative understanding has been
achieved as a first step.

\section*{2. Nonlinear dynamics}

The covariant Lagrangian, or 1+3 covariant, approach is based on a
physical choice of a 4-velocity vector field $u^a$. All variables
are in principle physically measurable by comoving observers. This
approach is inherently nonlinear. It starts from the inhomogeneous
and anisotropic universe, without a priori restrictions on the
degree of inhomogeneity and anisotropy, and then applies the
Friedmann limit when required. The basic theoretical ingredients
are: the covariant Lagrangian hydrodynamics of Ehlers and Ellis
\cite{ehl,ell}, and the perturbation theory of Hawking \cite{haw}
and Ellis and Bruni \cite{eb} which is derived from it; the
covariant Lagrangian approach to kinetic theory of Ellis et al.
\cite{ETMa}; and the covariant analysis of temperature anisotropies
introduced by Maartens et al. \cite{mes} and developed by Challinor
and Lasenby \cite{cl,c} and Gebbie et al. \cite{GE}.

The projected symmetric tracefree (PSTF) parts of vectors and
rank-2 tensors are
\[
V_{\la a\ra}=h_a{}^bV_b\,,~
S_{\la ab\ra }= \left\{h_{(a}{}^ch_{b)}{}^d-
{\ts{1\over3}}h^{cd}h_{ab}\right\}S_{cd}\,,
\]
where $h_{ab}=g_{ab}+u_au_b$ is the projector, with $g_{ab}$ the
spacetime metric. The skew part of a projected rank-2 tensor is
spatially dual to the projected vector
$S_a={1\over2}\ep_{abc}S^{[bc]}$, where $\ep_{abc}=\eta_{abcd}u^d$
is the projection of the spacetime alternating tensor. Any
projected rank-2 tensor has the irreducible decomposition
\[
S_{ab}=\case{1}/{3}Sh_{ab}+\ep_{abc}S^c+S_{\la ab\ra}\,,
\]
where $S=S_{cd}h^{cd}$ is the spatial trace.
A covariant vector product and its generalization to PSTF
rank-2 tensors are
\[
[V,W]_a=\ep_{abc}V^bW^c\,,~[S,Q]_a=\ep_{abc}S^b{}_dQ^{cd}\,.
\]
The covariant derivative $\nabla_a$ produces time and spatial
derivatives
\[
\dot{J}^{a\cdots}{}{}_{\cdots b}= u^c \nabla_c
J^{a\cdots}{}{}_{\cdots b}\,,~
\D_c J^{a\cdots}{}{}_{\cdots b} =
    h_c{}^d h^a{}_e\cdots h_b{}^f
    \nabla_d J^{e\cdots}{}{}_{\cdots f}\,.
\]
The projected derivative $\D_a$ further splits irreducibly into a
spatial divergence and curl \cite{maa}
\begin{eqnarray*}
&& \div V=\D^aV_a\,,~(\div S)_a=\D^bS_{ab}\,, \\
&& \c V_a=\ep_{abc}\D^bV^c\,,~ \c
S_{ab}=\ep_{cd(a}\D^cS_{b)}{}^d\,.
\end{eqnarray*}

Relative motion of comoving observers is encoded in the kinematic
quantities: the expansion $\Theta=\D^au_a$, the 4-acceleration
$A_a\equiv\dot{u}_a=A_{\la a\ra }$, the vorticity
$\omega_a=-{\ts{1\over2}}\c u_a$, and the shear
$\sigma_{ab}=\D_{\la a}u_{b\ra }$. The dynamic quantities describe
the sources of the gravitational field: the (total) energy density
$\rho=T_{ab}u^au^b$, isotropic pressure $p={1\over3}h_{ab}T^{ab}$,
energy flux $q_a=-T_{\la a\ra b}u^b$, and anisotropic stress
$\pi_{ab}=T_{\la ab\ra}$, where $T_{ab}$ is the total
energy-momentum tensor. The locally free gravitational field, i.e.
the part of the spacetime curvature not directly determined locally
by dynamic sources, is given by the Weyl tensor $C_{abcd}$. This
splits irreducibly into the gravito-electric and gravito-magnetic
fields
\[
E_{ab}=C_{acbd}u^cu^d=E_{\la ab\ra }\,,~~
H_{ab}={\ts{1\over2}}\ep_{acd}C^{cd}{}{}_{be}u^e=H_{\la ab\ra} \,,
\]
which provide a covariant Lagrangian description of tidal forces
and gravitational radiation.

The Ricci identity for $u^a$ and the Bianchi identities $\nabla^d
C_{abcd} = \nabla_{[a}(-R_{b]c} + {1\over6}Rg_{b]c})$ produce the
fundamental evolution and constraint equations governing the above
covariant quantities \cite{ehl,ell}. Einstein's equations are
incorporated via the algebraic replacement of the Ricci tensor
$R_{ab}$ by $T_{ab}-{1\over2}T_c{}^cg_{ab}$. These equations, in
fully nonlinear form and for a general source of the
gravitational field, are:

{\sf Evolution:}
\begin{eqnarray}
 \dot{\rho} +(\rho+p)\Theta+\div q=
  -2A^a q_a
 -\sigma^{ab}\pi_{ab}\,, && \label{e1}\\
\dot{\Theta} +{\ts{1\over3}}\Theta^2
+{\ts{1\over2}}(\rho+3p)-\div A =
 -\sigma_{ab}\sigma^{ab}&&\nonumber\\
{}+2\omega_a\omega^a+A_aA^a \,,&&
\label{e2}\\
 \dot{q}_{\la a\ra }
+{\ts{4\over3}}\Theta q_a+(\rho+p)A_a +\D_a p
+(\div\pi)_{a} =
-\sigma_{ab}q^b &&\nonumber\\
{}+[\omega,q]_a
-A^b\pi_{ab}\,, &&
\label{e3} \\
 \dot{\omega}_{\la a\ra } +{\ts{2\over3}}\Theta\omega_a
+{\ts{1\over2}}\c A_a =\sigma_{ab}\omega^b \,, && \label{e4}\\
 \dot{\sigma}_{\la ab\ra } +{\ts{2\over3}}\Theta\sigma_{ab}
+E_{ab}-{\ts{1\over2}}\pi_{ab} -\D_{\la a}A_{b\ra } =
-\sigma_{c\la a}\sigma_{b\ra }{}^c&&\nonumber\\
{}-
\omega_{\la a}\omega_{b\ra }
+A_{\la a}A_{b\ra }\,, &&
\label{e5}\\
 \dot{E}_{\la ab\ra } +\Theta E_{ab}
-\c H_{ab}
+{\ts{1\over2}}(\rho+p)\sigma_{ab}
+{\ts{1\over2}}
\dot{\pi}_{\la ab\ra }+{\ts{1\over6}}
\Theta\pi_{ab}&&\nonumber\\
{}+{\ts{1\over2}}\D_{\la a}q_{b\ra }
=-A_{\la a}q_{b\ra }
 +2A^c\ep_{cd(a}H_{b)}{}^d
+3\sigma_{c\la a}E_{b\ra }{}^c &&\nonumber\\
{}-\omega^c \ep_{cd(a}E_{b)}{}^d
-{\ts{1\over2}}\sigma^c{}_{\la a}\pi_{b\ra c}
-{\ts{1\over2}}\omega^c\ep_{cd(a}\pi_{b)}{}^d \,, &&
\label{e6}\\
\dot{H}_{\la ab\ra } +\Theta H_{ab}
+\c E_{ab}
-{\ts{1\over2}}\c\pi_{ab}=
3\sigma_{c\la a}H_{b\ra }{}^c-\omega^c \ep_{cd(a}H_{b)}{}^d
&&\nonumber\\
{}-2A^c\ep_{cd(a}E_{b)}{}^d
-{\ts{3\over2}}\omega_{\la a}q_{b\ra }+
{\ts{1\over2}}\sigma^c{}_{(a}\ep_{b)cd}q^d \,. &&
\label{e7}
\end{eqnarray}

{\sf Constraint:}
\begin{eqnarray}
\div\omega =A^a\omega_a \,, && \label{c1}\\
(\div\sigma)_{a}-\c\omega_a
-{\ts{2\over3}}\D_a\Theta
+q_a =-
2[\omega,A]_a \,, && \label{c2}\\
  \c\sigma_{ab}+\D_{\la a}\omega_{b\ra }
 -H_{ab}= -2A_{\la a}
\omega_{b\ra } \,, &&  \label{c3}\\
 (\div E)_{a}
+{\ts{1\over2}}(\div\pi)_{a}
 -{\ts{1\over3}}\D_a\rho
+{\ts{1\over3}}\Theta q_a
=[\sigma,H]_a &&\nonumber\\
{}-3H_{ab}\omega^b
+{\ts{1\over2}}\sigma_{ab}q^b-{\ts{3\over2}}
[\omega,q]_a \,, && \label{c4}\\
 (\div H)_{a}
+{\ts{1\over2}}\c q_a
 -(\rho+p)\omega_a =
-[\sigma,E]_a&&\nonumber\\
{}-{\ts{1\over2}}[\sigma,\pi]_a
+3E_{ab}\omega^b -{\ts{1\over2}}\pi_{ab}
\omega^b \,. && \label{c5}
\end{eqnarray}

If the universe is almost Friedmann, then quantities that vanish in
the Friedmann limit are $O(\epsilon)$, where $\epsilon$ is a
dimensionless smallness parameter, and the quantities are suitably
normalised (e.g. $\sqrt{\sigma_{ab}\sigma^{ab}}/\Theta<\epsilon$,
etc.). Linearisation reduces all the right hand sides of the
evolution and constraint equations to zero.

For a given choice of fundamental frame $u^a$, each of the species
$I$ which source the gravitational field has relative velocity
$v_{\!I}$ and 4-velocity
\be
u_{\!_I}^a=\gamma_{\!_I}\left(u^a+v_{\!_I}^a\right)\,,~v_{\!_I}^au_a=0
\,.\label{t3}
\ee
If the 4-velocities are close, i.e. if the frames are in
non-relativistic relative motion, then
$O(v^2)$ terms may be dropped from the equations,
except if we include nonlinear kinematic, dynamic and
gravito-electric/magnetic effects, in which case, for consistency, we
must retain $O(\epsilon^0v^2)$
terms such as $\rho v^2$, which are of the same order of magnitude
in general as $O(\epsilon^2)$ terms. 
This is the physically relevant nonlinear regime,
i.e. the case where only
nonrelativistic bulk velocities
are considered, but no
restrictions are imposed on non-velocity terms, and
we neglect only terms $O(\epsilon v^2,v^3)$.
Second-order effects involve neglecting also terms $O(\epsilon^3)$,
while linearisation means
neglecting terms $O(\epsilon^2,\epsilon v,v^2)$.

The dynamic quantities in the evolution and constraint
equations (\ref{e1})--(\ref{c5}) are the total quantities, with
contributions from all dynamically significant particle species.
Thus
\begin{eqnarray}
T^{ab} &=& \sum_I T_{\!_I}^{ab} = \rho u^au^b+ph^{ab}+2q^{(a}u^{b)}
+\pi^{ab} \,, \label{t1}\\
T_{\!_I}^{ab}&=& \rho_{\!_I}u_{\!_I}^au_{\!_I}^a+p_{\!_I}h_{\!_I}^{ab}
+2q_{\!_I}^{(a}u_{\!_I}^{b)}+\pi_{\!_I}^{ab}\,. \label{t2}
\end{eqnarray}
The dynamic quantities $\rho_{\!_I},\cdots$ in equation
(\ref{t2}) are as measured in the $I$-frame. For cold dark matter (CDM),
baryons, photons and neutrinos
\[
p_{\!_C}=0=q_{\!_C}^a=\pi_{\!_C}^{ab}\,,~~
q_{\!_B}^a=0=\pi_{\!_B}^{ab}\,,~~
p_{\!_R}=\case{1}/{3}\rho_{\!_R}\,,~~p_{\!_N}=\case{1}/{3}
\rho_{\!_N}\,,
\]
where we have chosen the unique 4-velocity in the CDM and
baryonic cases which follows from modelling these fluids as perfect.
The cosmological constant is characterized by
\[
p_{\!_V}=-\rho_{\!_V}=-\Lambda\,,~q_{\!_V}^a=0=\pi_{\!_V}^{ab}\,,~~
v_{\!_V}^a=0\,.
\]

The conservation equations for the species are best given in the
overall $u^a$-frame, as are the evolution and constraint equations
above. This requires the expressions for the partial dynamic
quantities as measured in the overall frame. The inverse velocity
relation is
\[
u^a=\gamma_{\!_I}\left(u_{\!_I}^a+v_{\!_I}^{*a}\right)\,,~~
v_{\!_I}^{*a}=-\gamma_{\!_I}\left(v_{\!_I}^a+v_{\!_I}^2u^a\right)\,,
\]
where $v_{\!_I}^{*a}u_{\!_Ia}=0$, and $v_{\!_I}^{*a}v_{\!_Ia}^*=
v_{\!_I}^av_{\!_Ia}$.
Then the dynamic quantities of species $I$ as
measured in the {\em overall $u^a$-frame} are
\begin{eqnarray}
\rho_{\!_I}^* &=& \rho_{\!_I}
+ \left\{\gamma_{\!_I}^2v_{\!_I}^2\left(\rho_{\!_I}+p_{\!_I}\right)
+2\gamma_{\!_I}q_{\!_I}^a
v_{\!_Ia}+\pi_{\!_I}^{ab}v_{\!_Ia}v_{\!_Ib}\right\} \,,\label{t4}\\
p_{\!_I}^* &=&  p_{\!_I}
+{\textstyle{1\over3}}
\left\{\gamma_{\!_I}^2v_{\!_I}^2\left(\rho_{\!_I}
+p_{\!_I}\right)+2\gamma_{\!_I}q_{\!_I}^a
v_{\!_Ia}+\pi_{\!_I}^{ab}v_{\!_Ia}v_{\!_Ib}\right\}\,, \label{t5}\\
q_{\!_I}^{*a} &=& q_{\!_I}^a+(\rho_{\!_I}+p_{\!_I})v_{\!_I}^a
\nonumber\\
&&{}+\left\{ (\gamma_{\!_I}-1)q_{\!_I}^a
-\gamma_{\!_I}q_{\!_I}^bv_{\!_Ib}u^a
+\gamma_{\!_I}^2v_{\!_I}^2
\left(\rho_{\!_I}+p_{\!_I}\right)v_{\!_I}^a\right.\nonumber\\
&&\left.{}+\pi_{\!_I}^{ab}v_{\!_Ib}-\pi_{\!_I}^{bc}v_{\!_Ib}v_{\!_Ic}u^a
\right\} \,,
\label{t6}\\
\pi_{\!_I}^{*ab} &=& \pi_{\!_I}^{ab} +
\left\{-2u^{(a}\pi_{\!_I}^{b)c}v_{\!_Ic}+\pi_{\!_I}^{bc}v_{\!_Ib}
v_{\!_Ic}u^au^b\right\} \nonumber\\
&&{}+\left\{-\case{1}/{3}\pi_{\!_I}^{cd}v_{\!_Ic}v_{\!_Id}h^{ab}+
\gamma_{\!_I}^2\left(\rho_{\!_I}+p_{\!_I}\right)
v_{\!_I}^{\la a}v_{\!_I}^{b\ra}+2\gamma_{\!_I}v_{\!_I}^{\la a}q_{\!_I}
^{b\ra}
\right\}\,.
\label{t7}
\end{eqnarray}
These are the nonlinear generalisations of well-known linearised
results, which correspond to removing all terms in braces,
dramatically simplifying the expressions. To linear order, there is
no difference in the dynamic quantities when measured in the
$I$-frame or the fundamental frame, apart from a simple velocity
correction to the energy flux. But in the general nonlinear case,
this is no longer true.

The total dynamic quantities are simply given by
\[
\rho=\sum\rho_{\!_I}^*\,,~p=\sum p_{\!_I}^*\,,~
q^{a}=\sum q_{\!_I}^{*a}\,,~\pi^{ab}=\sum\pi_{\!_I}^{*ab}\,.
\]
A convenient choice for each partial 4-velocity $u_{\!_I}^a$ is the
energy frame, i.e. $q_{\!_I}^a=0$ for each $I$. In the fundamental
frame, the partial energy fluxes do not vanish, i.e.
$q_{\!_I}^{*a}\neq0$, and the total energy flux is given by
\[
q^a=\sum\left[\left(\rho_{\!_I}+p_{\!_I}\right)v_{\!_I}^a+
\pi_{\!_I}^{ab}v_{\!_Ib}+O(\epsilon v_{\!_I}^2,v_{\!_I}^3) \right]\,.
\]
Then the dynamic quantities of CDM as measured in the fundamental
frame are
\begin{eqnarray}
&& \rho_{\!_C}^*=\gamma_{\!_C}^2\rho_{\!_C}\,,~~p_{\!_C}^*=
{\ts{1\over3}}
\gamma_{\!_C}^2v_{\!_C}^2\rho_{\!_C}\,,\label{t12}\\
&& q_{\!_C}^{*a}=\gamma_{\!_C}^2\rho_{\!_C}v_{\!_C}^a\,,~~
\pi_{\!_C}^{*ab}=\gamma_{\!_C}^2\rho_{\!_C}
v_{\!_C}^{\langle a}v_{\!_C}^{b\rangle}\,, \label{t13}
\end{eqnarray}
while for baryonic matter:
\begin{eqnarray}
&& \rho_{\!_B}^*=\gamma_{\!_B}^2\left(1+w_{\!_B}v_{\!_B}^2\right)
\rho_{\!_B}\,,
~~p_{\!_B}^*=\left[w_{\!_B}+{\ts{1\over3}}
\gamma_{\!_B}^2v_{\!_B}^2(1+w_{\!_B})\right]\rho_{\!_B}\,,
\label{t14}\\
&& q_{\!_B}^{*a}=\gamma_{\!_B}^2(1+w_{\!_B})\rho_{\!_B}v_{\!_B}^a\,,~~
\pi_{\!_B}^{*ab}=\gamma_{\!_B}^2(1+w_{\!_B})\rho_{\!_B}
v_{\!_B}^{\langle a}v_{\!_B}^{b\rangle}\,, \label{t15}
\end{eqnarray}
where $w_{\!_B}\equiv p_{\!_B}/\rho_{\!_B}$. For radiation and
neutrinos, the dynamic quantities relative to the $u^a$-frame are
found directly via kinetic theory below.

The total energy-momentum tensor is conserved, i.e.
$\nabla_bT^{ab}=0$, which is equivalent to the evolution
equations (\ref{e1}) and (\ref{e3}).
The partial energy-momentum tensors obey
\begin{equation}
\nabla_bT_{\!_I}^{ab}=J_{\!_I}^{a}=U_{\!_I}^*u^a+M_{\!_I}^{*a}\,,
\label{t16}
\end{equation}
where $U_{\!_I}^*$ is the rate of energy density transfer to
species $I$ as measured in the $u^a$-frame, and
$M_{\!_I}^{*a}=M_{\!_I}^{*\langle a\rangle}$ is the rate of
momentum density transfer. Thus
\[
J_{\!_C}^a=0=J_{\!_N}^a\,,~~
J_{\!_R}^a=-J_{\!_B}^a=U_{\!_T}u^a+M_{\!_T}^a\,,
\]
where the Thomson rates are, to $O(\epsilon\vb^2,\vb^3)$,
\begin{eqnarray*}
U_{\!_T}&=&\ne\st\left({\ts{4\over3}}\rho_{\!_R}^*\vb^2-q_{\!_R}^{*a}
v_{\!_Ba}\right)\,,\\
M_{\!_T}^a &=&\ne\st\left({\ts{4\over3}}\rho_{\!_R}^*\vb^a
-q_{\!_R}^{*a}+\pi_{\!_R}^{*ab}v_{\!_Bb}\right) \,,
\end{eqnarray*}
as shown below. Here $\ne$ is the free electron number density, and
$\st$ is the Thomson cross-section. Note that beyond linear order
there is energy transfer, i.e. $U_{\!_T}\neq 0$.

Using equations (\ref{t12})--(\ref{t15}) in (\ref{t16}), we find
that, to $O(\epsilon v^2,v^3)$, for CDM
\begin{eqnarray}
\dot{\rho}_{\!_C}+\Theta\rho_{\!_C}+\rho_{\!_C}\div v_{\!_C} =
-\left(\rho_{\!_C}v_{\!_C}^2\right)^{\rd}-{\ts{4\over3}}v_{\!_C}^2
\Theta \rho_{\!_C}&&\nonumber\\
{} -v_{\!_C}^a\D_a\rho_{\!_C}-2\rho_{\!_C}A_av_{\!_C}^a
 \,,&& \label{t19}\\
\dot{v}_{\!_C}^a+{\ts{1\over3}}\Theta v_{\!_C}^a+A^a =
A_bv_{\!_C}^bu^a-\sigma^a{}_bv_{\!_C}^b
+[\omega,v_{\!_C}]^a-v_{\!_C}^b\D_bv_{\!_C}^a
\,,&& \label{t20}
\end{eqnarray}
and for baryonic matter
\begin{eqnarray}
\dot{\rho}_{\!_B}+\Theta(1+w_{\!_B})\rho_{\!_B}+(1+w_{\!_B})
\rho_{\!_B}\div v_{\!_B}
=-\left[(1+w_{\!_B})\rho_{\!_B}v_{\!_B}^2\right]^{\rd}
&&\nonumber\\
{}-{\ts{4\over3}}v_{\!_B}^2\Theta(1+w_{\!_B})
\rho_{\!_B}
-v_{\!_B}^a\D_a\left[(1+w_{\!_B})\rho_{\!_B}\right]&&\nonumber\\
{}-2(1+w_{\!_B})\rho_{\!_B}A_av_{\!_B}^a
-\ne\st\left({\ts{4\over3}}\rho_{\!_R}^*\vb^2-
q_{\!_R}^{*a}v_{\!_Ba}\right)
\,,&& \label{t21}\\
(1+w_{\!_B})\dot{v}_{\!_B}^a+\left({\ts{1\over3}}-c_{\!_B}^2\right)
\Theta v_{\!_B}^a+(1+w_{\!_B})A^a
{}+\rho_{\!_B}^{-1}\D^ap_{\!_B}&&\nonumber\\
{}+\rho_{\!_B}^{-1}\ne\st
\left(\rho_{\!_R}^*\vb^a-q_{\!_R}^{*a}\right)
=
(1+w_{\!_B})A_bv_{\!_B}^bu^a&&\nonumber\\
{}-(1+w_{\!_B})\sigma^a{}_bv_{\!_B}^b
 +(1+w_{\!_B})[\omega,v_{\!_B}]^a
-(1+w_{\!_B})v_{\!_B}^b\D_bv_{\!_B}^a &&\nonumber\\{}+
 c_{\!_B}^2(1+w_{\!_B})(\div \vb)\vb^a
{}-\rho_{\!_B}^{-1}\ne\st\pi_{\!_R}^{*ab}v_{\!_Bb}
\,,&& \label{t22}
\end{eqnarray}
where $c_{\!_B}^2\equiv\dot{p}_{\!_B}/\dot{\rho}_{\!_B}$ (this
equals the adiabatic sound speed only to linear order). These are
the {\em nonlinear energy conservation and relative velocity
equations for matter.} Linearization reduces the right hand sides
to zero, dramatically simplifying the equations. The conservation
equations for the massless species (radiation and neutrinos) are
found below.

\section*{3. Nonlinear Thomson scattering}

In covariant Lagrangian kinetic theory \cite{ETMa} the photon
4-momentum is split as
\[
p^a=E(u^a+e^a)\,,~~e^a e_a=1\,,~e^a u_a=0\,,
\]
where $E=-u_ap^a$ is the energy and $e^a=p^{\la a\ra}/E$ is the
direction, as measured by a comoving (fundamental) observer. The
photon distribution function is expanded in covariant harmonics
\[
f(x,p)=f(x,E,e)
=F+F_ae^a+F_{ab}e^ae^b+\cdots\,= \sum_{\ell\geq0} F_{A_l}(x,E)
e^{\la A_l\ra},
\]
where ${e}^{A_\ell}\equiv e^{a_1}e^{a_2}\cdots e^{a_\ell}$
and
\[
F_{a\cdots b}=F_{\la a\cdots b\ra}~~\Leftrightarrow~~
F_{a\cdots b}=F_{(a\cdots b)}\,,~F_{a\cdots b}u^b=0=
F_{a\cdots bc}h^{bc}\,.
\]
The first 3 multipoles arise from the radiation energy-momentum
tensor,
\[
T_{\!_R}^{ab}(x)=\int p^ap^bf(x,p)\d^3p=
\rho_{\!_R}^*u^au^b+{\ts{1\over3}}\rho_{\!_R}^*h^{ab}
+2q_{\!_R}^{*(a}u^{b)}+\pi_{\!_R}^{*ab}\,.
\]
The radiation brightness multipoles are
\[
\Pi_{a_1\cdots a_\ell} =
\int E^3 F_{a_1\cdots a_\ell}\d E\,,
\]
so that (dropping asterisks) $\Pi=\rho_{\!_R}/4\pi$,
$\Pi^a=3q_{\!_R}^{a}/4\pi$ and $\Pi^{ab}=15\pi_{\!_R}^{ab}/8\pi$.
These multipoles define the temperature fluctuations \cite{mes}.

The Boltzmann equation is
\[
{\d f\over \d v}\equiv p^a{\p f\over \p x^a}-\Gamma^a{}_{bc}
p^bp^c{\p f\over \p p^a}=C[f]=b+b_ae^a+b_{ab}e^ae^b+\cdots
 \,,
\]
where the collision term $C[f]$ determines the rate of change of
$f$ due to emission, absorption and scattering processes. For
simplicity, the effects of polarization are neglected (see \cite{c}
for a linear treatment), and
\[
C[f]=\st \ne E_{\!_B}\left[\bar{f}(x,p)-f(x,p)\right]\,,
\]
where $E_{\!_B}=-p_a\ub^a$ is the photon energy relative to the
baryonic frame $\ub^a$, and \cite{cl}
\be
\bar{f}(x,p)={3\over 16\pi}\int f(x,p')\left[1+\left(
\eb^{a}e'_{\!_Ba}\right)^2\right]\d\Omega'_{\!_B} \,.
\label{r4}\ee
Here $e'_{\!_Ba}$ is the initial and $\eb^a$ is the final
direction, so that
\[
p'^a=E_{\!_B}\left(\ub^a+\eb'^a\right)\,,~
p^a=E_{\!_B}\left(\ub^a+\eb^a\right)\,,
\]
where we have used $E'_{\!_B}=E_{\!_B}$, which follows since the
scattering is elastic. The exact forms of the photon energy and
direction in the baryonic frame are
\begin{eqnarray*}
E_{\!_B} &=& E \gamma_{\!_B}\left(1-\vb^a e_a\right) \,, \\
\eb^a &=& {1\over \gamma_{\!_B}(1-\vb^ce_c)}\left[e^a+
\gamma^2_{\!_B}\left(\vb^be_b-\vb^2\right)u^a+\gamma^2_{\!_B}
\left(\vb^be_b-1\right)\vb^a\right]\,.
\end{eqnarray*}

Since the baryonic frame will move non-relativistically relative to
the fundamental frame in all cases of physical interest, it is
sufficient to linearize only in $\vb$, and not in the other
quantities. Thus we drop terms in $O(\epsilon\vb^2,\vb^3)$ but do
not neglect terms that are $O(\epsilon^0\vb^2,\epsilon\vb)$ or
$O(\epsilon^2)$ relative to the background. It follows from
equation (\ref{r4}) that
\[
4\pi\int \bar{f}E_{\!_B}^3\d E_{\!_B}=(\rho_{\!_R})_{\!_B}+
{\ts{3\over4}}(\pi_{\!_R}^{ab})_{\!_B}e_{\!_Ba}e_{\!_Bb} \,,
\]
where the dynamic radiation quantities are evaluated in the {\em
baryonic} frame. Transforming back to the fundamental frame, we
find, to $O(\epsilon\vb^2,\vb^3)$,
\begin{eqnarray*}
(\rho_{\!_R})_{\!_B} &=& \rho_{\!_R}\left[1+{\ts{4\over3}}\vb^2\right]
-2q_{\!_R}^a v_{\!_Ba} \,,\\
(\pi_{\!_R}^{ab})_{\!_B}& =& \pi_{\!_R}^{ab}+2v_{\!_Bc}
\pi_{\!_R}^{c(a}u^{b)}-2q_{\!_R}^{\la a}\vb^{b\ra}
+{\textstyle{4\over3}}\rho_{\!_R}\vb^{\la a}\vb^{b\ra}\,.
\end{eqnarray*}
Using the above equations and various identities \cite{mge}, and
defining the energy-integrated scattering multipoles
\[
K_{A_\ell}=\int E^2b_{A_\ell}\d E\,,
\]
we find that to $O(\epsilon\vb^2,\vb^3)$
\begin{eqnarray}
K &=& \ne\st\left[{\ts{4\over3}}\Pi\vb^2-
{\ts{1\over3}}\Pi^av_{\!_Ba}\right]\,,
\label{rr1}\\
K^a &=&-\ne\st\left[\Pi^a-4\Pi\vb^a-{\ts{2\over5}}\Pi^{ab}v_{\!_Bb}
\right]\,,
\label{rr2}\\
K^{ab} &=&-\ne\st\left[{\ts{9\over10}}\Pi^{ab}-{\ts{1\over2}}
\Pi^{\la a}\vb^{b\ra}-{\ts{3\over7}}\Pi^{abc}v_{\!_Bc}
-3\Pi\vb^{\la a}\vb^{b\ra}
\right]\,,
\label{rr3}\\
K^{abc} &=&-\ne\st\left[\Pi^{abc}-{\ts{3\over2}}
\Pi^{\la ab}\vb^{c\ra}-{\ts{4\over9}}\Pi^{abcd}v_{\!_Bd}
\right]\,,
\label{rr3a}
\end{eqnarray}
and, for $\ell> 3$:
\begin{equation}
K^{A_\ell}=
-\ne\st\left[
\Pi^{A_\ell}-\Pi^{\la A_{\ell-1}} \vb^{a_\ell\ra}
-\left({\ell+1\over 2\ell+3}\right)
\Pi^{A_\ell a}v_{\!_Ba}\right]\,.
\label{r21}
\end{equation}
Equations (\ref{rr1})--(\ref{r21}) are a nonlinear generalisation
of the linearised Thomson scattering results \cite{cl}. They show
clearly the {\em coupling of baryonic bulk velocity to the radiation
multipoles, arising from local nonlinear effects in Thomson
scattering.}

The multipoles of $E^{-1}\d f/\d v$ are derived in \cite{ETMa} and
\cite{mge}, using different methods. The result is
\begin{eqnarray*}
&&
\dot{F}_{\la A_\ell \ra}-{\ts{1\over3}}\Theta EF'_{A_\ell}
+\D_{\la a_\ell} F_{A_{\ell-1}\ra} +{(\ell+1)\over(2\ell+3)}
\D^aF_{aA_\ell}
\\
&&{}
-{(\ell+1)\over(2\ell+3)}E^{-(\ell+1)}\left[E^{\ell+2}F_{aA_\ell}
\right]'A^a-E^\ell\left[E^{1-\ell}F_{\la A_{\ell-1}}\right]'
A_{a_\ell\ra}
\\
&&{}
-\ell\omega^b\ep_{bc( a_\ell}F_{A_{\ell-1})}{}^c
-{(\ell+1)(\ell+2)\over(2\ell+3)(2\ell+5)}E^{-(\ell+2)}
\left[E^{\ell+3}F_{abA_\ell}\right]'\sigma^{ab}
\\
&&{}
-{2\ell\over (2\ell+3)}E^{-1/2}\left[E^{3/2}F_{b\la A_{\ell-1}}
\right]'\sigma_{a_\ell\ra}{}^b-E^{\ell-1}\left[E^{2-\ell}
F_{\la A_{\ell-2}}\right]'\sigma_{a_{\ell-1}a_\ell\ra}\,,
\end{eqnarray*}
where a prime denotes $\partial/\partial E$. This result is exact
and holds for any photon or (massless) neutrino distribution in any
spacetime. Integrating, it leads to
\begin{eqnarray}
K_{A_\ell} &=&
\dot{\Pi}_{\la A_\ell\ra}+{\ts{4\over3}}\Theta \Pi_{A_\ell}+
\D_{\la a_\ell}\Pi_{A_{\ell-1}\ra}
+{(\ell+1)\over(2\ell+3)}\D^b \Pi_{bA_\ell}
\nonumber\\
&&{}
-{(\ell+1)(\ell-2)\over(2\ell+3)} A^b \Pi_{bA_\ell}
+(\ell+3) A_{\la a_\ell} \Pi_{A_{\ell-1}\ra}
-\ell\omega^b\ep_{bc( a_\ell} \Pi_{A_{\ell-1})}{}^c
\nonumber\\
&&{}
-{(\ell-1)(\ell+1)(\ell+2)\over(2\ell+3)(2\ell+5)}
\sigma^{bc}\Pi_{bcA_\ell}+{5\ell\over(2\ell+3)}
\sigma^b{}_{\la a_\ell} \Pi_{A_{\ell-1}\ra b}\nonumber\\&&{}
-(\ell+2)
\sigma_{\la a_{\ell}a_{\ell-1}} \Pi_{A_{\ell-2}\ra}\,.
\label{r26}\end{eqnarray}
For decoupled neutrinos, $K_{\!_N}^{A_\ell}=0$, while for
(unpolarised) photons undergoing Thomson scattering, the left hand
side of equation (\ref{r26}) is given by equations
(\ref{rr1})--(\ref{r21}), which are exact in the kinematic and
dynamic quantities, but first order in the baryonic bulk velocity.
These equations show {\em the coupling of acceleration, vorticity
and shear to the radiation multipoles that arises at nonlinear
level.}

The monopole and dipole of equation (\ref{r26}) imply photon
conservation of energy and momentum density (to
$O(\epsilon\vb^2,\vb^3)$):
\begin{eqnarray}
\dot{\rho}_{\!_R}+{\ts{4\over3}}\Theta\rho_{\!_R}+\D_a q_{\!_R}^a
+2A_aq_{\!_R}^a+\sigma_{ab}\pi_{\!_R}^{ab}
=\ne\st\left({\ts{4\over3}}\rho_{\!_R}\vb^2-q_{\!_R}^av_{\!_Ba}
\right)\,,&&
\label{nl5}\\
\dot{q}_{\!_R}^{\la a\ra}+{\ts{4\over3}}\Theta q_{\!_R}^a
+{\ts{4\over3}}\rho_{\!_R}A^a+{\ts{1\over3}}\D^a\rho_{\!_R}
+\D_b\pi_{\!_R}^{ab}
+\sigma^a{}_bq_{\!_R}^b-[\omega,q_{\!_R}]^a+A_b\pi_{\!_R}^{ab}&&\nonumber\\
{}=\ne\st\left({\ts{4\over3}}\rho_{\!_R}\vb^a-q_{\!_R}^a+
\pi_{\!_R}^{ab}v_{\!_Bb}\right)\,.&&
\label{nl6}
\end{eqnarray}
The quadrupole evolution equation is
\begin{eqnarray}
&&\dot{\pi}_{\!_R}^{\la ab\ra}+{\ts{4\over3}}\Theta \pi_{\!_R}^{ab}
+{\ts{8\over15}}\rho_{\!_R}\sigma^{ab}+
{\ts{2\over5}}\D^{\la a}q_{\!_R}^{b\ra}
+{8\pi\over35}\D_c \Pi^{abc}
\nonumber\\
&&{}+2 A^{\la a} q_{\!_R}^{b\ra}
-2\omega^c\ep_{cd}{}{}^{(a} \pi_{\!_R}^{b) d}
+{\ts{2\over7}}\sigma_c{}^{\la a}\pi_{\!_R}^{b\ra c}
-{32\pi\over315}
\sigma_{cd} \Pi^{abcd}
\nonumber\\
{}&&{}=
-\ne\st\left[{\ts{9\over10}}\pi_{\!_R}^{ab}-{\ts{1\over5}}
q_{\!_R}^{\la a}\vb^{b\ra}-{8\pi\over35}\Pi^{abc}v_{\!_Bc}
-{\ts{2\over5}}\rho_{\!_R}\vb^{\la a}\vb^{b\ra}
\right]\,,
\label{nl8}
\end{eqnarray}
and the higher multipoles ($\ell>3$) evolve according to
\begin{eqnarray}
&&\dot{\Pi}^{\la A_\ell\ra}+{\ts{4\over3}}\Theta \Pi^{A_\ell}+
\D^{\la a_\ell}\Pi^{A_{\ell-1}\ra}
+{(\ell+1)\over(2\ell+3)}\D_b \Pi^{bA_\ell}
\nonumber\\
&&{}-{(\ell+1)(\ell-2)\over(2\ell+3)} A_b \Pi^{bA_\ell}
+(\ell+3) A^{\la a_\ell} \Pi^{A_{\ell-1}\ra}
-\ell\omega^b\ep_{bc}{}{}^{( a_\ell} \Pi^{A_{\ell-1}) c }
\nonumber\\
&&{}-{(\ell-1)(\ell+1)(\ell+2)\over(2\ell+3)(2\ell+5)}
\sigma_{bc}\Pi^{bcA_\ell}+{5\ell\over(2\ell+3)}
\sigma_b{}^{\la a_\ell} \Pi^{A_{\ell-1}\ra b}\nonumber\\&&{}
-(\ell+2)
\sigma^{\la a_{\ell}a_{\ell-1}} \Pi^{A_{\ell-2}\ra}
\nonumber\\
&&{}=
-\ne\st\left[
\Pi^{A_\ell}-\Pi^{\la A_{\ell-1}} \vb^{a_\ell\ra}
-\left({\ell+1\over 2\ell+3}\right)
\Pi^{A_\ell a}v_{\!_Ba}\right]\,.
\label{nl9}
\end{eqnarray}
For $\ell=3$, the second term in square brackets on the right of
equation (\ref{nl9}) must be multiplied by ${3\over2}$. The
temperature fluctuation multipoles are determined by the radiation
brightness multipoles $\Pi_{A_\ell}$ \cite{mge}.

These equations show in a transparent and covariant form precisely
which physical effects are directly responsible for the evolution
of CMB anisotropies in an inhomogeneous universe. They apply at
second and higher-order, and are readily specialised to the linear
case. They show how the matter generates anisotropies: directly
through interaction with the radiation, as encoded in the Thomson
scattering terms on the right of equations
(\ref{nl5})--(\ref{nl9}); and indirectly through the generation of
inhomogeneities in the gravitational field via the field equations
(\ref{e1})--(\ref{c5}) and the evolution equation (\ref{t22}) for
the baryonic velocity $\vb^a$. This in turn feeds back into the
multipole equations via the kinematic quantities, the baryonic
velocity $\vb^a$, and the spatial gradient $\D_a\rho_{\!_R}$ in the
dipole equation (\ref{nl6}). The coupling of the multipole
equations themselves provides an up and down cascade of effects,
shown in general by equation (\ref{nl9}). Power is transmitted to
the $\ell$-multipole by lower multipoles through the dominant
(linear) distortion term $\D^{\la a_{\ell}}
\Pi^{A_{\ell-1} \ra}$, as well as through nonlinear terms coupled
to the 4-acceleration ($A^{\la a_\ell}\Pi^{A_{\ell-1}\ra}$),
baryonic velocity ($\vb^{\la a_\ell}\Pi^{A_{\ell-1}\ra}$), and
shear ($\sigma^{\la a_\ell a_{\ell-1}}\Pi^{A_{\ell-2}\ra}$).
Simultaneously, power cascades down from higher multipoles through
the linear divergence term $(\div\Pi)^{A_\ell}$, and the nonlinear
terms coupled to $A^a$, $\vb^a$ and $\sigma^{ab}$. The vorticity
coupling does not transmit across multipole levels.

\section*{4. Conclusion}

A range of nonlinear effects is identified via a systematic
covariant analysis. These include the following.
\begin{itemize}
\item
{\em Nonlinear relative velocity} corrections, as exemplified in the
dynamic quantities in equations (\ref{t4})--(\ref{t7}) and in the
bulk velocity equations (\ref{t20}) and (\ref{t22}).
\item
{\em Nonlinear Thomson scattering} corrections affect the baryonic and
radiation conservation equations, entailing a coupling of the
baryonic bulk velocity $\vb^a$ to the radiation energy density,
momentum density and anisotropic stress. The evolution of the
radiation quadrupole $\pi_{\!_R}^{ab}$ also acquires nonlinear
Thomson corrections, which couple $\vb^a$ to the radiation dipole
$q_{\!_R}^a$ and octopole $\Pi^{abc}$. Linearization, by removing
these terms, has the effect of removing the nonlinear contribution
of the radiation multipoles $\Pi^{A_{\ell\pm1}}$ to the collision
multipole $K^{A_\ell}$.

\item
{\em Nonlinear kinematic} corrections introduce additional
acceleration and shear terms. Vorticity corrections are purely
nonlinear, i.e. a linear approach could give the false impression
that vorticity has no direct effect at all on the evolution of CMB
anisotropies. However, for very high $\ell$, i.e. on very small
angular scales, the nonlinear vorticity term could in principle be
non-negligible. The general evolution equation (\ref{nl9}) for the
radiation brightness multipoles $\Pi^{A_\ell}$ shows that {\em five
successive multipoles,} i.e. for $\ell-2$, $\cdots$, $\ell+2$, are
linked together in the nonlinear case. The
4-acceleration $A_a$ couples to the $\ell\pm 1$ multipoles, the
vorticity $\omega_a$ couples to the $\ell$ multipole, and the shear
$\sigma_{ab}$ couples to the $\ell\pm2$ and $\ell$ multipoles. All
of these couplings are nonlinear, except for $\ell=1$ in the case
of $A_a$, and $\ell=2$ in the case of $\sigma_{ab}$. These latter
couplings that survive linearisation are shown in the dipole
equation (\ref{nl6}) (i.e. $\rho_{\!_R}A^a$) and the quadrupole
equation (\ref{nl8}) (i.e. $\rho_{\!_R}\sigma^{ab}$). The latter
term drives Silk damping during the decoupling process. The
disappearance of most of the kinematic terms upon linearisation is
further reflected in the fact that the linearised equations link
only {\em three} successive moments, i.e. $\ell$, $\ell\pm1$.
\item
A crucial feature of the nonlinear kinematic terms is that some
of them scale like $\ell$ for large $\ell$, as already noted in the
case of vorticity. There are no purely linear terms with this
property, which has an important consequence, i.e. that {\em for
very high $\ell$ multipoles (corresponding to very small angular
scales in CMB observations), certain nonlinear terms may reach the
same order of magnitude as the linear contributions.}
(Note that
the same effect applies to the neutrino background.) The relevant
nonlinear terms in Eq. (\ref{nl9}) are (for $\ell\gg 1$):
\begin{eqnarray*}
&&{}-\ell \left[
{\ts{1 \over 4}} \sigma_{bc} \Pi^{bc A_{\ell} }
+ \sigma^{\la a_{\ell} a_{\ell-1}} \Pi^{A_{\ell-2}\ra}\right.\\
&&\left.{}+A^{\la a_\ell}\Pi^{A_{\ell-1}\ra}+{\ts{1\over2}}A_b\Pi^{bA_\ell}
+\omega^b\ep_{bc}{}{}^
{\la a_\ell}\Pi^{A_{\ell-1}\ra c}\right]\,.
\end{eqnarray*}
Any observable imprint of this effect will be made after last
scattering. In the free-streaming era, it is reasonable to
neglect the vorticity relative to the shear. We can remove the
acceleration term by choosing $u^a$ as the
dynamically dominant cold dark matter frame (i.e.
choosing $v_{\!_C}^a=0$).
It follows that the
nonlinear correction to the rate of change of the
linearised fluctuation multipoles is
\be
\delta(\dot{\Pi}^{A_\ell})\sim
\ell \left[
{\ts{1 \over 4}} \sigma_{bc} \Pi^{bc A_{\ell} }
+ \sigma^{\la a_{\ell} a_{\ell-1}} \Pi^{A_{\ell-2}\ra}\right]
~\mbox{ for }~\ell\gg1\,.
\label{r40}\ee
The linear solutions for $\Pi_{A_\ell}$ and $\sigma_{ab}$
can be used
in equation (\ref{r40}) to estimate the
correction to second order.
Its effect on observed anisotropies will be estimated
by integrating $\delta(\dot{\Pi}^{A_\ell})$
from last scattering to now.

\end{itemize}

~\\

Further quantitative analysis of the new nonlinear effects in
\cite{mm2,mge} is needed to identify more clearly which effects
could be observationally significant. This would also clarify the
relationship between various nonlinear effects that have been
derived under different assumptions and using different approaches
\cite{ov,sz2,rs,mm2,mge,hss}.

\[ \]
{\bf Acknowledgements}

It is a pleasure to acknowledge the inspiration provided by the
energy, enthusiasm and ideas of George Ellis, whose influence is
widely felt in relativistic cosmology.


\end{document}